%% file: main.tex
\documentclass{article}

\usepackage[utf8]{inputenc}
\usepackage{amsmath,amssymb,amsthm}
\usepackage{authblk}
\usepackage{comment}
\usepackage{hyperref}
\usepackage{graphicx}
\graphicspath{ {./figures/} }
\usepackage{caption}
\usepackage{subcaption}

\usepackage{algorithm}
\usepackage{algorithmic}

\algsetup{
    indent=2em,
    linenodelimiter=.
}
\floatname{algorithm}{Function}

\title{Random Formula Generators}

\author[1,2]{Ariel Jonathan Roffé}
\author[2,3]{Joaquín S. Toranzo Calderón}
\affil[1]{{\footnotesize Centro de Estudios de Filosofía e Historia de la Ciencia, Universidad Nacional de Quilmes-CONICET (CEFHIC-UNQ-CONICET), Argentina}}
\affil[2]{{\footnotesize Buenos Aires Logic Group (BA-Logic), SADAF, Argentina}}
\affil[3]{{\footnotesize Grupo de Inteligencia Artificial y Robótica (GIAR), Universidad Tecnológica Nacional, Argentina}}
\date{\today}

\begin{document}

\maketitle

\begin{abstract}
\noindent In this article, we provide three generators of propositional formulae for arbitrary languages, which uniformly sample three different formulae spaces. They take the same three parameters as input, namely, a desired depth, a set of atomics and a set of logical constants (with specified arities). The first generator returns formulae of \textit{exactly} the given depth, using \textit{all or some} of the propositional letters. The second does the same but samples \textit{up-to} the given depth. The third generator outputs formulae with \textit{exactly} the desired depth and \textit{all} the atomics in the set. To make the generators uniform (i.e. to make them return every formula in their space with the same probability), we will prove various cardinality results about those spaces.

\medskip

\noindent \textbf{Key-words:} Random Formula Generator, Formulae Set Cardinality, Finite Languages, Uniform Sampling

\end{abstract}

\section{Introduction}

This paper aims to provide algorithms for some random generators of propositional formulae. The motivation for this project was born out of a practical need. One of the authors of this work (Roffé) was designing software for logic teaching---which is now published as \textit{TAUT} \cite{roffe_TAUT}. The various modules of this software were intended to be capable of randomly generating logic exercises (as well as their solution and the correction of the input provided by the end-user), a problem that includes the random generation of formulae. Moreover, for the random formula generators to fulfill their function within \textit{TAUT}, there are several desirable requirements they should meet. These are:

\begin{enumerate}
\item In order for the end-user to be able to control the difficulty of the exercises, the generator should be able to provide formulae with---either \textit{up-to} or \textit{exactly}---a given number of different propositional letters and a given depth (for instance, in the truth table modules, the number of letters and depth control the number of columns and rows, respectively).

\item Since \textit{TAUT} works with many logic systems (some of which, such as LFI1 \cite{carnielli_coniglio_2016}, add new vocabulary) and in some modules (e.g. the truth table modules) the end-users can even define their own vocabulary and truth tables, it would be desirable for the generators to work for any finite propositional language, with any finite set of logical constants.

\item Given a particular depth and number of propositional letters, every formulae containing them should be returned with a probability greater than zero.

\item Every formula containing them should be generated with \textit{the same} probability (i.e. the generator should return formulae with a uniform distribution). Note that this requisite entails, and thus is stronger, than the previous one.
\end{enumerate}

\noindent To accomplish this, it would seem reasonable to just implement existing generators. However, this would not be the best choice, for the following reasons. Random formula generators have been developed and deployed mainly in relation to \textit{Automatic Theorem Provers} (ATPs) \cite{sutcliffe_TPTP, sutcliffe_sutner_1998}. At a certain point in their development, ATPs reached such a level of proficiency that it became difficult to compare them with manually generated problems \cite{biere_heule_vanmaaren_walsh_2009, buro_kleinebuning_1992}. The solution was to implement random generators, which followed some specific design rules aimed at generating ``hard" formulae, and thus allowing comparison of the ATPs in terms of their performance on solving them \cite{giunchiglia_sebastiani_1996b, mitchell_selman_levesque_1992}. The design rules were such that the generators were biased, generating formulae with certain fixed structures.\footnote{The current standard is to test ATPs by having them solve formulae with \textit{clause normal form} (CNF), or other similar structures, depending on the language one is working with. This allows researchers to select specific parameters for variation and performance comparison, such as atomics (or negations of atomics) per clause and the number of clauses. In addition, some explicit combinations of literals will be avoided, when they make the formula become trivial. \cite{biere_heule_vanmaaren_walsh_2009, jarvisalo_leberre_roussel_simon_2012}} In other words, formulae with other structures (those too simple to solve for an ATP) were not generated at all.\footnote{There is evidence showing that allowing every structural parameter to vary freely in the formula generators can lead to incorrect conclusions about the performance of the \textit{ATP}. This is due to the fact that, if that is allowed, the generators produce trivial (or easy to solve) formula structures. There has been work in designing biased generators, which produce formula structures that are harder to solve. \cite{biere_heule_vanmaaren_walsh_2009, goldberg_1979, hustadt_schmidt_1997, hustadt_schmidt_2002, mitchell_selman_levesque_1992}}

Given the aforementioned desired properties, these generators are not completely suitable. Firstly, because the end-user would not be able to control the difficulty of the exercises by controlling the number of different propositional letters and the depth of the formulae to be generated. Secondly, because these generators are designed for specific languages, mainly classical propositional logic and some modal extensions of it \cite{giunchiglia_sebastiani_1996b, giunchiglia_sebastiani_1996a, horrocks_1997, rintanen_1999}, this would not allow the end-users to define their own vocabulary. And thirdly, as said above, because some formulae would not even be generated at all, while those generated would not come out with the same probability. Thus, some new generators are required, following different design rules that fulfill the desired properties.

This article provides various generators (the $G$ functions). Section 2 begins with a first version of a generator that, given a set $P$ of propositional letters, returns formulae with a depth of exactly $n$ (given as input) and only (but possibly not all) members of $P$ ($G_{ES}$, for \textit{Exactly} $n$ and \textit{Some} $P$). We prove that this initial version satisfies properties 1-3 but not 4. In order to make it satisfy 4, we will need to prove some results about the cardinalities of certain sets of formulae (the $Q$ functions), which we do in section 3. Section 4 describes how those results are used to give a version of $G_{ES}$ that satisfies requisite 4. In section 5, we use the results from the previous sections to provide other generators, such as $G_{US}$ (up-to $n$ depth and some $P$) and $G_{EA}$ (exactly $n$ and all $P$). Finally, we draw some conclusions. As supplementary material, we provide a set of Python modules where the main algorithms of this paper are implemented.

\section{An Initial Algorithm}

In this section, we provide an initial version of $G_{ES}$, which returns every formula with positive (but not equal) probability. This generator will be a recursive function, taking as input: the desired depth $n$ of the formulae to be generated, a set $P$ of atomic formulae and a set $C$ of logical constants (with any given arities). The general strategy will be to build the formulas "top-down". That is, if the depth given as input is $d$, it will choose a logical constant and call itself recursively (with a depth less than $d$) to build the formulae that fit into its arguments. When the depth reaches 0, it returns an atomic.\footnote{Throughout the paper, for reasons of computational tractability and simplicity, we work with prefix notation. Therefore, we write $\neg(p)$ and $\wedge(p, p)$ instead of the more usual $\neg p$ and $p \wedge p$.} In more detail, the generator can be specified as follows.

\input{G_ES_biased.tex}

\noindent Note that steps 6 and 7 ensure that the entire formula reaches the desired depth for at least one of the arguments of $\star$. However, the rest of the arguments do not necessarily have to reach that depth (hence, in step 9, the first argument to be given to $G_{ES}$ in the following step is chosen at random between 0 and $n-1$). In steps 5, 6 and 9 (and in many places below in the paper) we assume that the programming language being used contains a function to uniformly sample the members of a set.

Thus, for example, if $n = 1$, $P = \{ p_1 \}$ and $C = \{ \wedge \}$, the following will happen at execution time: 

\begin{itemize}
    \item Initial depth is not 0, so the execution enters the condition on step 4
    \item In step 5, the binary constant $\wedge$ is chosen
    \item In step 6, either 1 or 2 is chosen (let's suppose it is 2)
    \item In step 7, $G_{ES}(0, \{ p_1 \}, \{ \wedge \})$ is called
    \begin{itemize}
        \item Initial depth is 0, so the second execution enters the condition on step 1
        \item In step 2, $p_1$ is chosen
        \item In step 3, $p_1$ is returned to the top level execution
    \end{itemize}
    \item $s_2$ is assigned $p_1$
    \item In step 9, integer 0 is chosen from the set $\{ 0 \}$
    \item In step 10, $G_{ES}(0, \{ p_1 \}, \{ \wedge \})$ is called (as before, it returns $p_1$)
    \item $s_1$ is assigned $p_1$
    \item In step 12, $\wedge(p_1, p_1)$ is returned
\end{itemize}

\noindent Note that, by design, $G_{ES}$ ---as presented above--- will satisfy requisites 1 and 2 stated in the introduction. We now prove that it satisfies requisite 3.\\

\noindent \textbf{Theorem 1.} For all $n$, all $P$ and all $C$, $G_{ES}(n, P, C)$ returns every possible formula of depth $n$ and some $P$ with positive probability.

\begin{proof}
We prove this result by induction on the depth $n$ (we take $P$ and $C$ to be arbitrary finite sets of atomic formulae and logical constants during the whole proof).

If $n=0$, then $G_{ES}$ returns an atomic chosen at random from $P$. Thus, every formula of depth 0 has positive probability of being returned by $G_{ES}(0, P, C)$

For the inductive case, suppose that all formulae with depth $j<n$ are returned with positive probability. Let $f$ be an arbitrary formula (of form $\star(f_1, \dots, f_m)$). We show that $f$ has a positive probability of being returned in step 11. This follows immediately from the following two facts:

\begin{itemize}
\item The constant $\star$ with which $f$ begins has a positive probability of being sampled in step 5, in the first recursive iteration of $G_{ES}(n, P, C)$ (since $C$ is finite and $\star$ is a member of $C$).
\item Each of the $f_1, \dots, f_m$ is generated through recursive calls to $G_{ES}$, either in step 7 or 10. Whichever the step, $G_{ES}$ will be called with a depth lower than $n$. Thus, by the inductive hypothesis, each $f_i$ has a positive probability of being returned.
\end{itemize}

\end{proof}

\noindent Next, we show that, even though every formula has a positive probability of being returned by $G_{ES}$, it is not true that every formula has \textit{the same} probability of being returned. That is, we show that the version of $G_{ES}$ introduced above does not sample its formula space uniformly. The reasons why this is so will be illustrative of the kinds of problems we will have to solve to make it uniform.

Consider a case where $n = 1$, $P = \{ p_1, p_2 \}$ and $C = \{ \neg, \wedge \}$. $G_{ES}(n, P, C)$ will return formulae with either an $\neg(\varphi)$ or an $\wedge(\varphi, \varphi)$ form. Moreover, since in step 5 the constant is chosen uniformly, those two structures will appear with the same probability (0.5). However, $\neg(\varphi)$ has two instances---$\neg(p_1)$ and $\neg(p_2)$---while $\wedge(\varphi, \varphi)$ has four instances---$\wedge(p_1, p_1)$, $\wedge(p_1, p_2)$, $\wedge(p_2, p_1)$ and $\wedge(p_2, p_2)$. Thus, each of the negated formulas will be returned with probability $0.5 \cdot 0.5 = 0.25$, while each of the conjunctive ones will have probability $0.5 \cdot 0.25 = 0.125$ (figure 1).

\begin{figure}[H]
\centering
\includegraphics[width=0.8\textwidth]{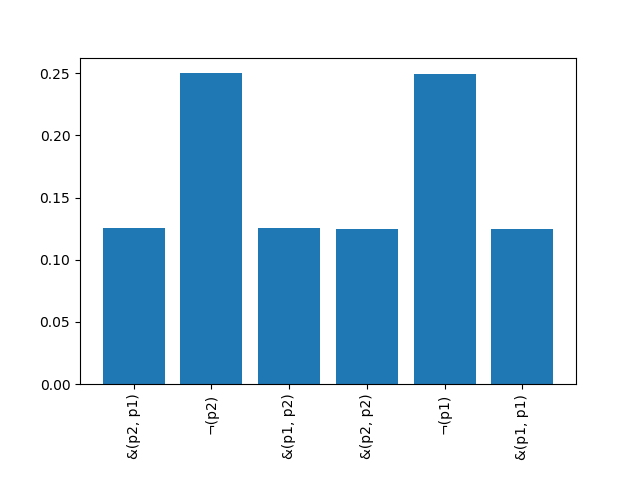}
\caption{Frequencies of the formulae obtained for $n = 1$, $P = \{ p_1, p_2 \}$ and $C = \{ \neg, \wedge \}$ for 1,000,000 generated formulae. Generated with the G\_ES\_biased function in the supplementary Python modules.}
\end{figure}

\noindent A second kind of case would be one where $n = 2$, $P = \{ p_1 \}$ and $C = \{ \wedge \}$. Now three structures are possible---$\wedge(\wedge(\varphi, \varphi), \varphi)$, $\wedge(\varphi, \wedge(\varphi, \varphi))$ and $\wedge(\wedge(\varphi, \varphi)$, $\wedge(\varphi, \varphi))$. A formula with the first structure will be outputted with probability 0.25 (the left conjunct is chosen with probability 0.5 in step 6, and depth 0 is chosen with probability 0.5 in step 9). The same happens with the second structure. The third, however, will be outputted with probability 0.5, since it may emerge in two different ways (left disjunct is chosen in step 6, depth 1 is chosen in step 9; and right disjunct is chosen in step 6, and depth 1 is chosen in step 9). Thus, given that there is only one propositional letter to choose from, $\wedge(\wedge(p_1, p_1), p_1)$ and $\wedge(p_1, \wedge(p_1, p_1))$ will have probability 0.25 each, while $\wedge(\wedge(p_1, p_1), \wedge(p_1, p_1))$ will have probability 0.5 (figure 2). 

\begin{figure}[H]
\centering
\includegraphics[width=0.6\textwidth]{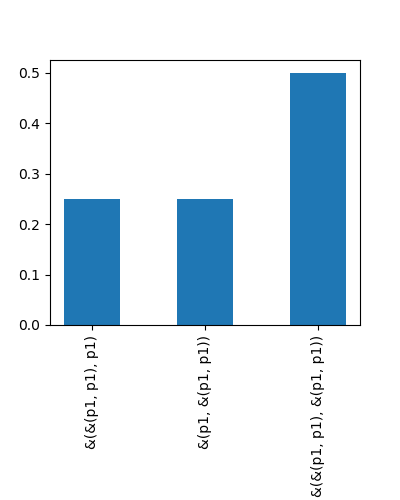}
\caption{Frequencies of the formulae obtained for $n = 2$, $P = \{ p_1 \}$ and $C = \{ \wedge \}$ for 1,000,000 generated formulae. Generated with the G\_ES\_biased function in the supplementary Python modules.}
\end{figure}

\noindent In contrast, if $P$ contained three propositional letters, each formula that is an instance of the first two structures would have a greater probability of being generated than a formula with the third structure (the calculations are similar, readers can perform them on their own).

To avoid both kinds of issues, we will need to weigh both the probabilities of choosing constants of given arities in step 5, as well as the subformula depths chosen in step 9. To do that, we will present equations which allow us to calculate various results about the cardinalities of certain sets of formulae, which we do in the following section.

\section{Counting Formulae}

In order to obtain a generator with a uniform distribution, we will present an algorithm that weighs the various parameters during structure generation. We will accomplish this by counting the number of formulae with (up-to or exactly) a given depth, including (all or some) members of a given set of atomic formulae, and a given set of logical constants. The present section introduces three equations that serve to count the cardinalities of those sets of formulae. For each, we will prove that they quantify what is intended.

The first function (which we call $Q_{US}$) counts how many formulae there are with up-to $n$ depth, including only (and all or some) propositional letters of a finite set $P$.\\

\noindent \textbf{Theorem 2.} For a language $\mathcal{L}$ with logical constant set $C$, and given a finite set of propositional letters $P$ (with $p = |P|$), the number of formulae with up-to $n$ depth that include all or some formulae in $P$ is given by:

\[ Q_{US}(n, p, C) =
    \begin{cases} 
      p & \hspace{5mm} \text{if } n = 0 \\
      p + \sum\limits_{i=1}^k C(i) \cdot Q_{US}(n-1, p, C)^{i} & \hspace{5mm} \text{if } n > 0 \\
   \end{cases}
\]

\begin{proof}
The proof is by induction on the depth $n$. For $n=0$ (the base case), the result is obvious. For the inductive hypothesis, assume that for all $m < n$, $\#_1(m, P, C) = Q_{US}(m, p, C)$ (where $\#_1(m, P, C)$ is the number of formulae with up-to $n$ depth and including all or some P).

A formula with depth at most $n$ is either an atomic formula (there are $p$ of those) or has a logical constant as its main symbol. We obtain the result by summing over the number of formulae that have each of the constants as the main symbol. That is,

\[ 
    \#_1(n, P, C) = p + \sum\limits_{i=1}^k C(i) \cdot \#_2(i, n, P, C)
\]

\noindent where $k = max(i)$ such that $C(i) \neq 0$ ($k$ is the maximum arity of the constants in $C$) and $\#_2(i, n, P, C)$ is the number of formulae with at most $n$ depth and some $P$ that begin with an (arbitrary) $i$-ary connective.

A formula with depth at most $n$ and $i$-ary connective $\star$ as its main symbol has the form $\star(f_1, ..., f_i)$. Since each of the $f_i$'s has a depth of at most $n-1$, by the inductive hypothesis, there are $Q_{US}(n-1, p, C)$ formulae that can fit into each of the arguments of $\star$. Hence, $\#_2(i, n, P, C) = Q_{US}(n-1, p, C)^{i}$, which immediately gives us the desired result.

\end{proof}

\noindent Having $Q_{US}$ allows us to easily define another equation, $Q_{ES}$, which counts the number of formulae with exactly $n$ depth that contain all or some members of a set $P$ of atomics. We do that by obtaining the number of formulae of up-to $n$ depth and subtracting the number of up-to $n-1$. We express this result as a corollary:\\

\noindent \textbf{Corollary 1.} For a language $\mathcal{L}$ with logical constant set $C$, the number of formulae with \textit{exactly} $n$ depth and some formulae from $P$ (such that $p$ = $|P|$) is:

\[ Q_{ES}(n, p, C) =
    \begin{cases} 
      p & \hspace{5mm} \text{if } n = 0 \\
      Q_{US}(n, p, C) - Q_{US}(n-1, p, C) & \hspace{5mm} \text{if } n > 0 \\
   \end{cases}
\]

\noindent It might be thought that $Q_{EA}(n, P, C)$ (exactly $n$ and all $P$) is easily definable in a similar way, by doing $Q_{ES}(n, p, C) - Q_{ES}(n, p-1, C)$. However, this is not the case. Consider for example a case with $n=1$, $P= \{ p_1, p_2, p_3 \}$ and $C= \{ \oplus \}$, where $\oplus$ is a 4-ary constant. $Q_{ES}(n, p, C)$ will count formulae such as $\oplus(p_1, p_1, p_1, p_1)$, $\oplus(p_2, p_2, p_2, p_2)$ and $\oplus(p_3, p_3, p_3, p_3)$, which we do not want to count (since they do not contain all members of $P$). However, subtracting $Q_{ES}(n, p-1, C)$ will only make us stop counting two of those formulae, not all three of them. That is because $Q_{ES}(n, p-1, C)$ counts the number of formulae with the same setting as above, but with $P= \{ p_1, p_2 \}$ (or, equivalently, removing one of the other two), which does not count $\oplus(p_3, p_3, p_3, p_3)$. Moreover, for the same reason, subtracting $Q_{ES}(n, p-1, C)$ will make us stop counting $\oplus(p_1, p_1, p_2, p_2)$ but not $\oplus(p_1, p_1, p_3, p_3)$ nor $\oplus(p_2, p_2, p_3, p_3)$ (or any other formulae that contains only two letters, $p_1$ and $p_3$, or $p_2$ and $p_3$).

To obtain $Q_{EA}(n, P, C)$ we first prove a preliminary proposition, which will be important in the first case of $Q_{EA}$.\\

\noindent \textbf{Proposition 2.} For a language $\mathcal{L}$ with logical constants of maximum arity $k$, a formula of depth $n$ can contain at most $k^n$ different propositional letters.\\

\begin{proof}
The proof is, once again, by induction on the depth $n$. 

For the base case, if $n = 0$, then $k^0 = 1$, which is obviously correct since a formula of depth 0 is a propositional letter (which, of course, contains at most one propositional letter).

For the inductive case, suppose that for all depths $j<n$, a formula may contain at most $k^j$ different propositional letters. What we need to calculate is the maximum number of propositional letters that a formula of depth $n$ will have.

Now, a formula of depth $n$ will have form $\star(f_1, \dots, f_m)$. Since, in this proposition, we are not imposing any restriction on the number of atomics that the formulae may contain, we can assume that $f_1, \dots, f_m$ do not share any propositional letters, in order to maximize the letters contained by $\star(f_1, \dots, f_m)$. Thus, the number of letters in $f_1, \dots, f_m$ will be given by $l(f_1) + \dots + l(f_m)$ (where $l(x)$ is the number of letters that $x$ has). 

Note also that each of the $f_1, \dots, f_m$ has a depth between 0 and $n-1$, and thus, by the inductive hypothesis, a number of letters between $k^0$ and $k^{n-1}$. Again, this number will be maximized when each $f_i$ has depth $n-1$. Thus, we get that $\star(f_1, \dots, f_m)$ will contain at most $m \cdot k^{n-1}$ letters. The only question left is the arity of $\star$ (and thus, what $m$ is). Again, this number will be maximized when $\star$ is a constant of the maximum arity available, which is $k$. Thus, $\star(f_1, \dots, f_m)$ will have at most $k \cdot k^{n-1} = k^{n}$ propositional letters.

\end{proof}

\noindent With this in mind, we can now provide an equation for $Q_{EA}$ as follows:\\

\noindent \textbf{Theorem 3.} For a language $\mathcal{L}$ with logical constant set $C$ (where $k = max(i)$ such that $C(i) \neq 0$; that is, $k$ is the maximum arity of the constants in $C$), if $P$ is a finite set of propositional letters (such that $p$ = $|P|$), the number of formulae with exactly $n$ depth and all propositional letters from $P$ is:

\[ Q_{EA}(n, p, C) =
    \begin{cases} 
      0 & \hspace{5mm} \text{if } p > k^n\\
      Q_{ES}(n, p, C) - \sum\limits_{j=i}^{p-1} \binom{p}{i} \cdot Q_{EA}(n, i, C) & \hspace{5mm} \text{otherwise}\\
   \end{cases}
\]

\begin{proof}
This time, the proof is by induction on the number of propositional letters in $P$, i.e. $p$.

For the base case ($p=1$), we must be in the second case of $Q_{EA}$, because whatever the values of $k$ and $n$ are, since $k$ must be a positive integer and $n$ will be equal to or greater than zero, it will not be the case that $1 > k^n$. Thus, since $P$ contains only one member, the number of formulae with exactly depth $n$ and some (but possibly not all) atomics in $P$ is equal to the number of formulae of exactly $n$ and all atomics in $P$. Thus, $Q_{EA}(n, p, C) = Q_{ES}(n, p, C)$. $Q_{EA}$ guarantees this since, in the second term of the second case, the sum will have 1 as lower bound and 0 as upper bound (and thus be equal to zero).

For the inductive hypothesis, suppose that for all $j<p$, $\#_3(n, j, C) = Q_{EA}(n, j, C)$, where $\#_3(n, j, C)$ is the number of formulae with exactly depth $n$ and containing all members of a set $J$ of formulae (with $j = |J|$).

Since now $p>1$, we will either be in case one (if $p > k^n$), for which the proof that there are 0 formulae is given in proposition 2, or we will be in case two (if $p \leq k^n$). From here on, we assume this is the case.

Now, since $p>1$, $Q_{ES}(n, p, C)$ will overcount, as it will include formulae that do not contain all members of $P$. The key is to subtract from it the number of formulae that contain $1, \dots, p-1$ atomics in $P$. Thus, if $\#_4(i, n, P, C)$ represents the number of formulae with exactly depth $n$ and including exactly $i$ elements of a set $P$, we get that:

\[ 
    Q_{EA}(n, p, C) = Q_{ES}(n, p, C) - \sum\limits_{i=1}^{p-1} \#_4(i, n, P, C)
\]

\noindent Next, note that $\#_4(i, n, P, C) = \binom{p}{i} \cdot \#_3(n, i, C)$. This is because there are $\binom{p}{i}$ ways of obtaining a set $I \subseteq P$ such that $|I| = i$. Thus, we get that:

\[ 
    Q_{EA}(n, p, C) = Q_{ES}(n, p, C) - \sum\limits_{i=1}^{p-1} \binom{p}{i} \cdot \#_3(n, i, C)
\]

\noindent The inductive hypothesis then directly gives us the desired result.

\end{proof}

\noindent In the next sections, we use these results to give non-biased versions of the random formula generators. However, we wish to stress that they are interesting on their own, and very likely have other uses besides the one given to them in the following sections.

\section{A Revised Algorithm}

In this section, we offer a revised version of $G_{ES}$ that samples every formula in its formula space with equal probability. We offer a proof of this last fact. In the next section, we use this algorithm to give other non-biased formula generators.

The uniform version of $G_{ES}$ can be specified by modifying Function 1 (i.e. the biased version of $G_{ES}$ presented above) in steps 5 and 9. In the case of step 5 (the choice of a logical constant), as noted at the end of section 2, the issue is that formulae have a constant of greater arity as their main symbol have a structure with more instances than those that have constants of a lesser arity. To briefly recapitulate, if $n = 1$ and $P = \{ p_1, p_2 \}$, $\neg(\varphi)$ has two instances while $\wedge(\varphi, \varphi)$ has four. Thus, we need the choice of logical constant to be weighted by the number of formulae that can be generated with that depth and that constant as a main symbol.

If there were no requirement that one of the arguments of the constant must reach depth $n-1$, then the number of formulae of depth $n$ that begin with an $m$-ary logical constant from $C$, and contain some $P$, would be given by $Q_{US}(n-1, |P|, C)^m$. However, since in $G_{ES}$ one of the arguments of the structure being generated must reach depth $n-1$, we must subtract the number of formulae that have depth $n-2$ or less in all their positions, which is given by $Q_{US}(n-2, |P|, C)^m$. Thus, step 4 will weigh the choice of logical constant in the following way. An $m$-ary constant will be chosen with probability:

\[ \frac{Q_{US}(n-1, |P|, C)^m - Q_{US}(n-2, |P|, C)^m}{Q_{ES}(n, |P|, C)} \]

\noindent The second problem with the previous version of $G_{ES}$ was located in step 9. Once a logical constant had been chosen, different choices of depth for its arguments also had different numbers of formulae as instances. For instance, suppose that $n = 2$, $P = \{ p_1 \}$ and $C=\{ \wedge \}$, Then, as explained above (see the end of Section 2), $\wedge(\wedge(\varphi, \varphi), \varphi)$ would be chosen with probability 0.25, while $\wedge(\wedge(\varphi, \varphi), \wedge(\varphi, \varphi))$ would have probability 0.5. Since each of those structures only has one instance, the corresponding formulae would also be returned with different probabilities.

Thus, again, we must weigh the choices of depth in step 9. One way to do so is the following. Suppose an $m$-ary constant $\star$ has been chosen in step 5. Then, the formula being generated will have the form $\star(f_1, \dots, f_m)$, where $f_1, \dots, f_m$ have depths $d_1, \dots, d_m$. For a particular choice of $d_1, \dots, d_m$, there will be $Q_{ES}(d_1, P, C) \cdot ... \cdot Q_{ES}(d_m, P, C)$ combinations of formulae that can be generated as the arguments of $\star$. Therefore, the distribution of depths $d_1, \dots, d_m$ (where at least one $d_i$ is equal to $n-1$) will be weighed with probability:

\[  \frac{Q_{ES}(d_1, P, C) \cdot ... \cdot Q_{ES}(d_m, P, C)}
{Q_{US}(n-1, |P|, C)^m - Q_{US}(n-2, |P|, C)^m} \]

\noindent The denominator in this equation is the same as the numerator above, which gives the total number of formulae of depth $n$ that start with an $m$-ary constant.

With these two modifications in mind, the uniform version of $G_{ES}$ can be given as follows:

\input{G_ES_uniform.tex}

\noindent As the reader can see in Figure 3, the implementation of this algorithm in the supplementary Python modules samples formulae with a uniform distribution.\\

\begin{figure}[H]
\centering
\begin{subfigure}{.6\textwidth}
  \centering
  \includegraphics[width=1\linewidth]{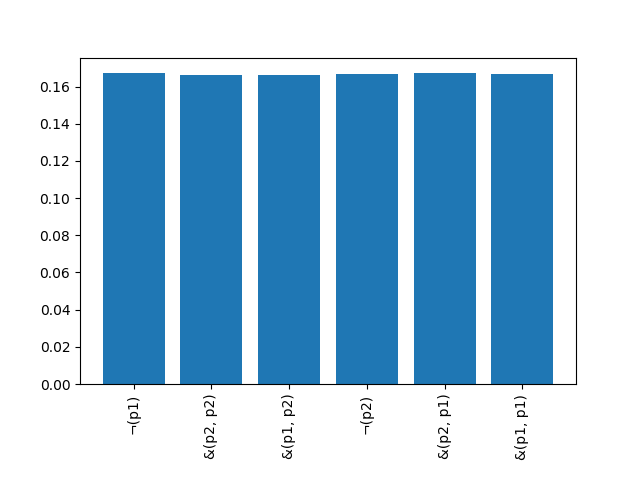}
  \caption{$n = 1$, $P = \{ p_1, p_2 \}$ and $C = \{ \neg, \wedge \}$}
\end{subfigure}%
\begin{subfigure}{.4\textwidth}
  \centering
  \includegraphics[width=1\linewidth]{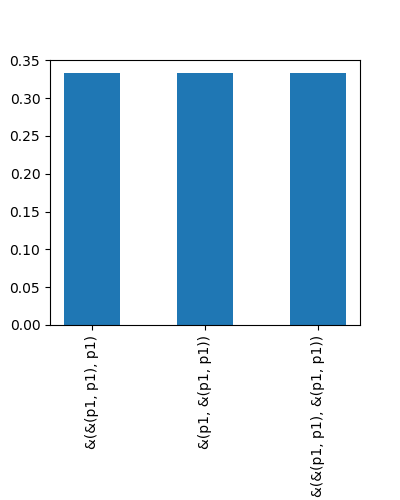}
  \caption{$n = 2$, $P = \{ p_1 \}$ and $C = \{ \wedge \}$}
\end{subfigure}
\caption{The same two cases from above. 1,000,000 formulae for each graph, generated with the G\_ES\_uniform function in the supplementary Python modules}
\end{figure}

\noindent \textbf{Theorem 4.}  For all $n$, all $P$ and all $C$, $G_{US}(n, E, C)$ returns every possible formula, of depth exactly $n$ and containing all or some members of $P$, with the same probability ($1/Q_{ES}(n, P, C)$).

\begin{proof}
The proof is by induction on the depth $n$. For the base case, when $n=0$, there are $Q_{ES}(0, P, C) = |P|$ possible formulae to be returned. The execution of $G_{ES}(0, P, C)$ will enter the condition on step 1 in its first recursive iteration, so the generator will return a uniformly sampled atomic from $P$. Thus, each possible outcome of $G_{ES}(0, P, C)$ will be returned with probability $1 / |P| = 1 / Q_{ES}(n, P, C)$

For the inductive hypothesis, suppose that every formula of depth $j < n$ is returned by $G_{ES}(j, P, C)$ with probability $1 / Q_{ES}(j, P, C)$.

The strategy here is to take an arbitrary formula $\star(f_1, \dots, f_m)$ of depth $n$ and show that it will be returned with probability $1 / Q_{ES}(n, P, C)$. For this purpose, consider the events that must take place during the execution of $G_{ES}$ so that $\star(f_1, \dots, f_m)$ is generated as output:

\begin{itemize}
    \item Constant $\star$ must be chosen in step 5.
    \item The distribution of depths $d_1, \dots, d_m$ (denoting the depths of $f_1, \dots, f_m$) must be chosen in step 6
    \item $f_1, \dots, f_m$ must be obtained in step 8 when $G_{ES}(d_1, P, C) \dots G_{ES}(d_m, P, C)$ are called
\end{itemize}

\noindent To get the probability that these three events occur, we must multiply their respective probabilities. The first two events will occur with probability:

\[ \frac{Q_{US}(n-1, |P|, C)^m - Q_{US}(n-2, |P|, C)^m}{Q_{ES}(n, |P|, C)} \cdot
\frac{Q_{ES}(d_1, P, C) \cdot ... \cdot Q_{ES}(d_m, P, C)}
{Q_{US}(n-1, |P|, C)^m - Q_{US}(n-2, |P|, C)^m}
\]

\[ = \frac{Q_{ES}(d_1, P, C) \cdot ... \cdot Q_{ES}(d_m, P, C)}{Q_{ES}(n, |P|, C)} \]

\medskip
\noindent Since $d_1, \dots, d_m$ are all lower than $n$, by the inductive hypothesis, the calls to $G_{ES}(d_1, P, C) \dots G_{ES}(d_m, P, C)$ will return $f_1, \dots, f_m$ with probability 

\[ \frac{1}{Q_{ES}(d_1, P, C)} \cdot ... \cdot  \frac{1}{Q_{ES}(d_m, P, C)} \]
\medskip

\noindent Thus, the probability of obtaining $\star(f_1, \dots, f_m)$ is given by:

\[\frac{Q_{ES}(d_1, P, C) \cdot ... \cdot Q_{ES}(d_m, P, C)}{Q_{ES}(n, |P|, C)} \cdot 
\frac{1}{Q_{ES}(d_1, P, C)} \cdot ... \cdot  \frac{1}{Q_{ES}(d_m, P, C)}\]

\[ = \frac{1}{Q_{ES}(n, |P|, C)} \]

\end{proof}

\noindent In the next section, we extend the results obtained in this section and in the previous one to provide two alternative uniform generators, $G_{US}$ and $G_{EA}$.

\section{Further algorithms}

In this section, we provide two further formula generators, which sample their respective spaces uniformly. The first is $G_{US}$, which returns a formula with depth up-to $n$ and some (but possibly not all) atomics in $P$.

$G_{US}$ is fairly easy to obtain from $G_{ES}$. For $G_{US}(n, P, C)$, we simply need to choose a depth $d$ from $\{ 0, \dots, n \}$ and then call $G_{ES}(d, P, C)$. The choice of depth, however, must be weighed, since there are different numbers of formulae for different depths. The way to do this is to choose a depth $d$ with probability:

\[ \frac{Q_{ES}(d, |P|, C)}{Q_{US}(n, |P|, C)} \]

\medskip
\noindent Thus, $G_{US}$ can be specified as follows:

\input{G_US.tex}

\noindent \textbf{Theorem 5.}  For all $n$, all $P$ and all $C$, $G_{US}(n, P, C)$ returns every possible formula, of depth up-to $n$ and containing all or some members of $P$, with the same probability ($1 / G_{US}(n, P, C)$).

\begin{proof}
Consider an arbitrary formula $f$ of depth $d$ (such that $0 \leq d \leq n$). For this formula to be generated by $G_{US}(n, P, C)$, $d$ must be chosen in step 1 and $f$ must be obtained from the call to $G_{ES}(d, P, C)$ in step 2. The weighing equation introduced above and Theorem 4 imply that this will occur with probability:

\[ \frac{Q_{ES}(d, |P|, C)}{Q_{US}(n, |P|, C)} \cdot 
\frac{1}{Q_{ES}(d, |P|, C)}
= \frac{1}{Q_{US}(d, |P|, C)}
\]

\end{proof}

\noindent Obtaining the next generator, $G_{EA}$, presents some additional complications. Firstly, if in step 2 the atomics are chosen from the entire set $P$, then some members of $P$ may never be chosen. For instance, if $n=1$, $P=\{ p_1, p_2 \}$ and $C=\{ \wedge \}$, then the formula we get will have the form $\wedge(\varphi, \varphi)$; but if in step 2 in both recursive paths $p_1$ is chosen, then the formula will not contain every atomic in $P$.

Secondly, if $n = 2$, $P = \{ p_1, p_2, p_3 \}$ and $C = \{ \neg, \wedge \}$, then the structure $\neg(\neg(\varphi))$ should never be chosen, since no instance of that structure can contain all three propositional letters. Likewise, if $n = 2$, $P = \{ p_1, p_2, p_3, p_4 \}$ and $C = \{ \wedge \}$, once the constant $\wedge$ has been chosen, the only depth distribution chosen with positive probability should be $(1, 1)$, since both $(0, 1)$ and $(1, 0)$ cannot accommodate 4 atomics. It seems, then, that the generator should somehow keep track of the number of atomics left to be placed, to ensure that every atomic appears in the final formula.

To accomplish that, the recursive step of the algorithm will not only choose a logical constant and a distribution of depths, but also a distribution of subsets of $P$ to assign to the arguments of the constant. That is, we will first generate a set $D$ that contains, for every every arity $m$ present in $C$, all possible distributions of form $((d_1, P_1), \dots, (d_m, P_m))$, where $d_i$ is the depth of the $i$-th argument and $P_i$ is the subset of $P$ for the $i$-th argument. For each argument of the main connective, $G_{EA}(d_i, P_i, C)$ will be called recursively. For instance, if $n=1$, $P = \{ p_1, p_2 \}$ and $C = \{ \neg, \wedge, \vee \}$, then $D$ will contain:

\begin{itemize}
    \item $((0, \{ p_1 \}), (0, \{ p_2 \}))$. If this structure is chosen, since it is of length 2, the formula will either be $\wedge(p_1, p_2)$ or $\vee(p_1, p_2)$ 
    \item $((0, \{ p_2 \}), (0, \{ p_1 \}))$. If this structure is chosen, since it is of length 2, the formula will either be $\wedge(p_2, p_1)$ or $\vee(p_2, p_1)$ 
    \item $(0, \{ p_1, p_2 \})$. This structure is of length 1, so we would choose $\neg$ as the main symbol, and for its argument recursively call $G_{EA}(0, \{p_1, p_2\}, C)$. However, this last part would fail (no formula of depth 0 can contain two atomics). Thus, this structure will be assigned weight 0 in the weighing function. This is how $G_{EA}$ keeps track of the atomics left to be placed.
\end{itemize}

\noindent More generally, $D$ will be such that:

\begin{itemize}
    \item Each member of $D$ has form $((d_1, P_1), \dots, (d_m, P_m))$, such that there is an $m$-ary constant in $C$
    \item For each member of $D$, there is at least one $d_x$ such that $d_x = n-1$
    \item For each member of $D$, each $P_i$ is a subset of $P$, and is not empty.
    \item For each member of $D$, $\bigcup_{x \in \{1, \dots, m\}} P_x = P$
\end{itemize}

\noindent To weigh a distribution of this sort, we must count how many formulae there are with exactly depth $d_x$ and all $P_x$ for all $x$ in $\{1, \dots, m\}$ (i.e. how many subformulae there are for each of the arguments of the main logical constant), over the total number of formulae possible. This will be given by:

\[ C(m) \cdot \frac{Q_{EA}(d_1, |P_1|, C) \cdot ... \cdot Q_{EA}(d_m, |P_m|, C)}{Q_{EA}(n, |P|, C)} \]

\noindent The $C(m)$ at the beginning is because a formula with distribution $((d_1, P_1), \dots, (d_m, P_m))$ can begin with any of the $m$-ary constants. With this in mind, the uniform version of $G_{EA}$ can be given as follows:

\input{G_EA.tex}

\noindent Notice that, if the condition on 1 is not fulfilled in the first recursive iteration, it will never be the case later on, since any distribution that contains $(d_i, P_i)$ as a member, where $|P_i| > k^{d_i}$, will be assigned a weight of 0.

\begin{figure}[H]
\centering
\includegraphics[width=0.8\textwidth]{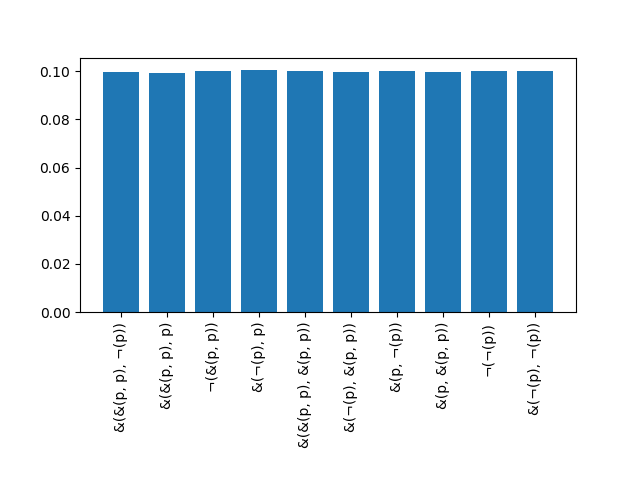}
\caption{Frequencies of the formulae obtained for $n = 2$, $P = \{ p \}$ and $C = \{ \neg, \wedge \}$ for 1,000,000 generated formulae. Generated with the G\_EA\_uniform function in the supplementary Python modules.}
\end{figure}

\noindent \textbf{Theorem 6.}  For all $n$, all $P$ and all $C$, $G_{EA}(n, P, C)$ returns every possible formula, of depth exactly $n$ that contains all members of $P$, with the same probability ($1 / Q_{EA}(n, P, C)$).

\begin{proof}
The proof is, once more, by induction on $n$. The base case is obvious. If $n=0$ and $|P|=1$, then $G_{EA}(0, P, C)$ will return the only possible formula with probability 1.

For the inductive case, suppose that for all $j < n$, $G_{EA}(j, P, C)$ returns every possible formula with probability $1 / Q_{EA}(j, |P|, C)$. We now show that an arbitrary formula $\star(f_1, \dots, f_m)$ has probability $1 / Q_{EA}(n, |P|, C)$ of being returned.

Three events have to take place for $\star(f_1, \dots, f_m)$ to be returned. The first is that the distribution $((d_1, P_1), \dots, (d_m, P_m))$ is chosen in step 6, where $d_1$ is the depth of $f_1$ and $P_1$ is the set of atomics in $f_1$, ..., and $d_m$ is the depth of $f_m$ and $P_m$ is the set of atomics in $f_m$. The second is that the formulae $f_1, \dots, f_m$ are generated through the recursive calls to $G_{EA}(d_1, P_1, C), \dots, G_{EA}(d_m, P_m, C)$ in step 8. The third is that constant $\star$ is chosen among the $m$-ary constants. The first probability is given by the weighing function, while the second by the inductive hypothesis, and their multiplication is equal to:

\[ C(m) \cdot \frac{Q_{EA}(d_1, |P_1|, C) \cdot ... \cdot Q_{EA}(d_m, |P_m|, C)}{Q_{EA}(n, |P|, C)} \cdot
\frac{1}{Q_{EA}(d_1, |P_1|, C)} \cdot ... \cdot \frac{1}{Q_{EA}(d_m, |P_m|, C)}
\]

\[ = \frac{C(m)}{Q_{EA}(n, |P|, C)} \]

Since, in step 10, the constant of arity $m$ is chosen uniformly, $\star$ will be chosen with probability $1/C(m)$, which directly gives us the desired result.

\end{proof}

\section{Conclusions}

In this article, we have introduced three random generators of propositional formulae, which we have called $G\_{ES}$, $G\_{US}$ and $G\_{EA}$. We began with a non-uniform version of $G\_{ES}$, proving that it returns every formula with positive, but not equal, probability. In section 3, results about the cardinalities of the formula spaces of those generators (named $Q\_{ES}$, $Q\_{US}$ and $Q\_{EA}$) were proven. This allowed us to give uniform versions of all three generators in sections 4 ($G\_{ES}$) and 5 (the other two).

There are two additional noteworthy features about the apparatus presented here. The first is that formulae containing sentential constants (such as $\top$ and $\bot$) can be generated with the above algorithms simply by treating them as extra atomics. The second is that the above generators can also be used as term generators in a first-order language with function symbols. For that purpose, the variables and individual constants must be treated as the atomics and the function symbols as the logical constants.

Having uniform formula generators can be useful for a variety of reasons. The one that led us here, as stated in the introduction, is related to the need of random formulae in the generation of logic exercises, but researchers will probably be able to find many more.

\bibliographystyle{abbrv}
\bibliography{bibliography}

\end{document}

%% file: G_ES_biased.tex
\begin{algorithm}[H]  
\caption{$G_{ES}(n, P, C)$ [Non-uniform version]}
\label{alg1}

\begin{algorithmic}[1]  

\REQUIRE $n$ (depth), $P$ (set of atomics), $C$ (logical constant set)
\ENSURE Random formula of depth $n$ using all or some atomics in $P$
    \IF {$n = 0$}
        \STATE Randomly choose an atomic $a$ from $P$
        \RETURN $a$
    \ELSE
        \STATE Randomly choose an $m$-ary connective $\star$ from $C$
        \STATE Randomly choose an integer $i$ from $\{ 1, \dots, m\}$
        \STATE $f_i \leftarrow G_{ES}(n-1, P, C)$
        \FORALL {$x$ in $\{ 1, \dots, m \} - i$}
            \STATE Randomly choose an integer $j$ from $\{ 0, \dots, n-1 \}$
            \STATE $f_x \leftarrow G_{ES}(j, P, C)$
        \ENDFOR
        \RETURN $\star(f_1, \dots, f_m)$
    \ENDIF

\end{algorithmic}
\end{algorithm}

%% file: G_ES_uniform.tex
\begin{algorithm}[H]
\caption{$G_{ES}(n, P, C)$ [Uniform version]}
\label{alg2}

\begin{algorithmic}[1]

\REQUIRE $n$ (depth), $P$ (set of atomics), $C$ (logical constant set)
\ENSURE Random formula of depth $n$ using all or some atomics in $P$
    \IF {$n = 0$}
        \STATE Randomly choose an atomic $a$ from $P$
        \RETURN $a$
    \ELSE
        \STATE Choose an $m$-ary connective $\star$ from $C$ with probability: 
        
        \[ \frac{Q_{US}(n-1, |P|, C)^m - Q_{US}(n-2, |P|, C)^m}{Q_{ES}(n, |P|, C)} \]
        
        \STATE Choose a distribution of depths $d_1, \dots, d_m$ (such that at least one $d_i = n-1$) with probability:
        
        \[  \frac{Q_{ES}(d_1, P, C) \cdot ... \cdot Q_{ES}(d_m, P, C)}
        {Q_{US}(n-1, |P|, C)^m - Q_{US}(n-2, |P|, C)^m} \]
        
        \FORALL {$x$ in $\{ 1, \dots, m \}$}
            \STATE $f_x \leftarrow G_{ES}(d_x, P, C)$
        \ENDFOR
        \RETURN $\star(f_1, \dots, f_m)$
    \ENDIF

\end{algorithmic}
\end{algorithm}

%% file: G_US.tex
\begin{algorithm}[H]
\caption{$G_{US}(n, P, C)$}
\label{alg3}

\begin{algorithmic}[1]

\REQUIRE $n$ (depth), $P$ (set of atomic formulae to use), $C$ (logical constant set)
\ENSURE Random formula

    \STATE Choose a depth $d$ from $\{0, \dots, n\}$ with probability $\frac{Q_{ES}(d, |P|, C)}{Q_{US}(n, |P|, C)}$
    \STATE $f \leftarrow G\_ES(d, P, C)$
    \RETURN $f$

\end{algorithmic}
\end{algorithm}

%% file: G_EA.tex
\begin{algorithm}[H]
\caption{$G_{EA}(n, P, C)$}
\label{alg4}

\begin{algorithmic}[1]

\REQUIRE $n$ (depth), $P$ (set of atomics), $C$ (logical constant set)
\ENSURE Random formula structure

    \IF {$|P| > k^n$, where $k$ is the maximum arity in $C$}
        \RETURN ERROR
    \ELSIF{$n = 0$}
        \RETURN the only member of $P$
    \ELSE
        \STATE Choose a distribution $((d_1, P_1), \dots, (d_m, P_m))$ from $D$ with probability:
        
        \[ C(m) \cdot \frac{Q_{EA}(d_1, |P_1|, C) \cdot ... \cdot Q_{EA}(d_m, |P_m|, C)}{Q_{EA}(n, |P|, C)} \]
        
        \FORALL {$x$ in $\{ 1, \dots, m \}$}
            \STATE $f_x \leftarrow G_{EA}(d_x, P_x, C)$
        \ENDFOR
        \STATE Randomly choose an $m$-ary constant $\star$ from C
        \RETURN $\star(f_1, \dots, f_m)$
    \ENDIF

\end{algorithmic}
\end{algorithm}

%% file: main.bbl
\begin{thebibliography}{10}

\bibitem{biere_heule_vanmaaren_walsh_2009}
A.~Biere, M.~Heule, H.~van Maaren, and T.~Walsh, editors.
\newblock {\em Handbook of Satisfiability}, volume 185 of {\em Frontiers in
  Artificial Intelligence and Applications}. IOS Press, 2009.

\bibitem{buro_kleinebuning_1992}
M.~Buro and H.~Kleine~B{\"u}ning.
\newblock {\em Report on a SAT competition. Technical Report 110}.
\newblock Department of Mathematics and Informatics, Universität Paderborn,
  Germany, November 1992.

\bibitem{carnielli_coniglio_2016}
W.~Carnielli and M.~Coniglio.
\newblock {\em Paraconsistent Logic: Consistency, Contradiction and Negation},
  volume~40 of {\em Logic, Epistemology, and the Unity of Science}.
\newblock Springer International Publishing, 06 2016.

\bibitem{giunchiglia_sebastiani_1996b}
F.~Giunchiglia and R.~Sebastiani.
\newblock Building decision procedures for modal logics from propositional
  decision procedures --- the case study of modal k.
\newblock In M.~A. McRobbie and J.~K. Slaney, editors, {\em Automated Deduction
  --- Cade-13}, pages 583--597, Berlin, Heidelberg, 1996. Springer Berlin
  Heidelberg.

\bibitem{giunchiglia_sebastiani_1996a}
F.~Giunchiglia and R.~Sebastiani.
\newblock A \textbf{SAT}-based decision procedure for \textit{ALC}.
\newblock In {\em Proceedings of the Fifth International Conference on
  Principles of Knowledge Representation and Reasoning}, KR'96, pages 304--314,
  San Francisco, CA, USA, 1996. Morgan Kaufmann Publishers Inc.

\bibitem{goldberg_1979}
A.~T. Goldberg.
\newblock {\em On the complexity of the satisfiability problem. Courant
  Computer Science Report 16}.
\newblock Computer Science Department, New York University, USA, October 1979.

\bibitem{horrocks_1997}
I.~Horrocks.
\newblock {\em Optimisation Techniques for Expressive Description Logics.
  Technical Report Series UMCS-97-2-1}.
\newblock Department of Computer Science, University of Manchester, UK,
  February 1997.

\bibitem{hustadt_schmidt_1997}
U.~Hustadt and R.~A. Schmidt.
\newblock On evaluating decision procedures for modal logic.
\newblock In M.~Pollack, editor, {\em Proceedings of the 15th International
  Joint Conference on Artifical Intelligence}, volume~1 of {\em IJCAI'97},
  pages 202--207, San Francisco, CA, USA, August 1997. Morgan Kaufmann
  Publishers Inc.

\bibitem{hustadt_schmidt_2002}
U.~Hustadt and R.~A. Schmidt.
\newblock Scientific benchmarking with temporal logic decision procedures.
\newblock In D.~Fensel, F.~Giunchiglia, D.~McGuinness, and M.~A. Williams,
  editors, {\em Principles of Knowledge Representation and Reasoning:
  Proceedings of the Eighth International Conference}, KR 2002, pages 533--546,
  San Francisco, CA, USA, April 2002. Morgan Kaufmann Publishers Inc.

\bibitem{jarvisalo_leberre_roussel_simon_2012}
M.~J{\"a}rvisalo, D.~Le~Berre, O.~Roussel, and L.~Simon.
\newblock The international \textbf{SAT} solver competitions.
\newblock {\em AI Magazine}, 33(1):89--94, March 2012.

\bibitem{mitchell_selman_levesque_1992}
D.~Mitchell, B.~Selman, and H.~Levesque.
\newblock Hard and easy distributions of \textbf{SAT} problems.
\newblock In {\em Proceedings of the Tenth National Conference on Artificial
  Intelligence}, AAAI'92, pages 459--465. AAAI Press, July 1992.

\bibitem{rintanen_1999}
J.~Rintanen.
\newblock Improvements to the evaluation of quantified boolean formulae.
\newblock In {\em Proceedings of the 16th International Joint Conference on
  Artificial Intelligence}, volume~2 of {\em IJCAI'99}, pages 1192--1197, San
  Francisco, CA, USA, October 1999. Morgan Kaufmann Publishers Inc.

\bibitem{roffe_TAUT}
A.~Roffé.
\newblock {TAUT}.
\newblock \url{https://www.taut-logic.com/}, 2018.
\newblock Accessed: 2019-09-17.

\bibitem{sutcliffe_TPTP}
G.~Sutcliffe.
\newblock The \textbf{TPTP} problem library and associated infrastructure. from
  \textbf{CNF} to \textbf{TH0}, \textit{TPTP} v6.4.0.
\newblock {\em Journal of Automated Reasoning}, 59(4):483--502, December 2017.

\bibitem{sutcliffe_sutner_1998}
G.~Sutcliffe and C.~Suttner.
\newblock The \textbf{TPTP} problem library.
\newblock {\em Journal of Automated Reasoning}, 21(2):177--203, October 1998.

\end{thebibliography}
